\documentclass[aps,prd,twocolumn,superscriptaddress,longbibliography,nofootinbib]{revtex4-2}
\usepackage{graphicx,longtable,mathrsfs,color,array}
\usepackage[hidelinks]{hyperref}
\usepackage{amsmath}
\usepackage[dvipsnames]{xcolor}
\graphicspath{{figures/}}
\usepackage{mathrsfs}
\usepackage{amssymb}
\usepackage{orcidlink}
\usepackage{comment}
\usepackage[normalem]{ulem}

\def\rcos{r_{\rm cos}}
\def\rcosprop{R_{\rm cos}}
\def\rprop{R}
\def\rod{r_{\rm od}}
\def\rodprop{R_{\rm od}}

\begin{document}

\title{Restrictions on Initial Conditions in Cosmological Scenarios and Implications for Simulations of Primordial Black Holes and Inflation}

\author{Thomas W.~Baumgarte\orcidlink{0000-0002-6316-602X}}
\email{tbaumgar@bowdoin.edu}
\affiliation{Department of Physics and Astronomy, Bowdoin College, Brunswick, Maine 04011, U.S.A.}

\author{Katy Clough\orcidlink{0000-0001-8841-1522}}
\email{k.clough@qmul.ac.uk}
\affiliation{Geometry, Analysis and Gravitation, School of Mathematical Sciences, Queen Mary University of London,
Mile End Road, London E1 4NS, United Kingdom}

\author{John T. Giblin, Jr.\orcidlink{0000-0003-1505-8670}}
\email{giblinj@kenyon.edu}
\affiliation{Department of Physics, Kenyon College, Gambier, Ohio 43022, U.S.A.}
\affiliation{CERCA/ISO, Department of Physics, Case Western Reserve University, Cleveland, Ohio 44106, U.S.A.}
\affiliation{Center for Cosmology and AstroParticle Physics (CCAPP) and Department of Physics, The Ohio State University, Columbus, Ohio 43210, U.S.A.}

\begin{abstract}
Numerical relativity simulations provide a means by which to study the evolution and end point of strong over-densities in cosmological spacetimes.  Specific applications include studies of primordial black hole formation and the robustness of inflation.  Here we adopt a toy model previously used in asymptotically flat spacetimes to show that, for given values of the over-density and the mean curvature, solutions to the Hamiltonian constraint need not exist, and if they do exist they are not unique.  Specifically, pairs of solutions exist on two branches, corresponding to strong-field and weak-field solutions, that join at a maximum beyond which solutions cease to exist. As a result, there is a limit to the extent to which an over-density can be balanced by intrinsic rather than extrinsic curvature on the initial slice.  Even below this limit, iterative methods to construct initial data may converge to solutions on either one of the two branches, depending on the starting guess, leading to potentially inconsistent physical results in the evolution.
\end{abstract}

\maketitle

\section{Introduction}

The evolution of the universe is governed by Einstein's equations of general relativity.  A homogeneous and isotropic universe is described {\em exactly} by a Friedmann-Lema\^itre-Robertson-Walker (FLRW) spacetime, which provides a fully non-linear solution to Einstein's equations.  While small inhomogeneities can be treated using a perturbative approach, i.e.~by linearizing Einstein's equations about an FLRW background, larger deviations from homogeneity will be affected by the nonlinear terms in Einstein's equations and hence require a non-perturbative treatment.  

Specific examples of cosmological processes that cannot be studied within the framework of perturbation theory include the possible formation of primordial black holes \cite{Carr:1974nx,Carr:1975qj,Misner:1964je,May:1966zz} (PBHs) and the robustness of inflationary models to inhomogeneities in the early universe.  Understanding whether, how, and where PBHs form \cite{1980Novikov,1979Bicknell,Hernandez:1966zia,1995Baumgarte,Niemeyer:1999ak,Shibata:1999zs,Green:2004wb,Hawke:2002rf,Musco:2004ak,1989Miller,Miller:1994xy,Polnarev:2006aa,Musco:2008hv,Musco:2012au,Saini:2017tsz,Musco:2018rwt,Musco:2021sva,Bloomfield:2015ila,Deng:2016vzb,Deng:2017uwc,Atal:2019erb,Escriva:2019nsa,Escriva:2019phb,Escriva:2021pmf,Escriva:2022bwe,Escriva:2022pnz,Escriva:2022yaf,Escriva:2023qnq,Harada:2023ffo,Yoo:2020lmg,Escriva:2024lmm} provides insights into the fundamental physics of inflation \cite{Marsh:2015xka,Garcia-Bellido:2017mdw,Ballesteros:2017fsr,Germani:2018jgr}, the nature of dark matter \cite{Ivanov:1994pa,Carr:2016drx,Carr:2020gox,Carr:2020xqk,Escriva:2022duf}, and the possible  origin of supermassive black holes \cite{Duechting:2004dk,Clesse:2015wea}.  Similarly, whether or not inflation, in the presence of severe early-time inhomogeneity, remains an attractor \cite{Brandenberger:2016uzh,Linde:2017pwt}
has also been investigated non-linearly \cite{Goldwirth:1989pr,East:2015ggf,Clough:2017efm,Aurrekoetxea:2019fhr,Joana:2020rxm,Corman:2022alv,Garfinkle:2023vzf,Joana:2024ltg,Elley:2024alx}, with one thorny question being what form of ``generic'' initial data is the right one to test such a scenario.
Since most of our cosmological intuition is anchored to the FLRW background, there is a tendency to imagine the non-linear physics exists on top of the underlying cosmology. However, the non-linear nature of Einstein's equations does not generally allow for such a separation.  In order to study these and other dynamic processes in cosmology, we have to solve the equations---including those for any initial conditions---self-consistently, and this may result in a solution very far from the FLRW one onto which any perturbations were added.

Numerical relativity has emerged as an important tool for the study of non-linear cosmological inhomogeneities (see, e.g., \cite{Aurrekoetxea:2024ypv} for a recent review).  Typically, numerical relativity simulations adopt a ``3+1" decomposition, in which the spacetime is foliated by spatial slices of constant coordinate time (see, e.g., \cite{Baumgarte:2010ndz} for a textbook treatment).  In such a 3+1 split, the spacetime curvature arises from a combination of curvature {\em intrinsic} to each slice, expressed by the spatial metric, and curvature {\em extrinsic} to each slice -- fittingly referred to as the extrinsic curvature -- which is related to the time derivative of the spatial metric.  

In practice, a numerical relativity simulation starts with initial data chosen on some spatial slice of constant coordinate time.  Note that there is an inherent contradiction in this construction - usually one does not know the exact 4D solution to the Einstein equation for the physical scenario (that is why it is being solved for numerically), and yet one wants to start with a 3D slice of this ``correct'' 4D solution. One aspect of this problem is that we need to construct initial data that solve the constraint equations, i.e.~the Hamiltonian and momentum constraints.  
Different approaches to this problem have been suggested \cite{Shibata:1999zs,Garfinkle:2020iup,Bentivegna:2013xna,Aurrekoetxea:2022mpw}, but 
it is common to express the matter (in particular the mass-energy density) and gravitational-field variables (the spatial metric and extrinsic curvature) as those of an FLRW background plus some deviation that need not be small.  Since the background variables solve the constraints identically, this amounts to solving the constraint equations for the (non-linear) deviations from the FLRW background.  The Hamiltonian constraint, in particular, can be interpreted as stating that an inhomogeneity in the mass-energy density must be balanced by inhomogeneities in either the intrinsic or the extrinsic curvature of the initial slice, or a combination of both.  Even though one might naively believe that we could freely choose which curvature, intrinsic or extrinsic, a given mass-energy deviation is balanced by, we show in this paper that this is {\em not} the case.

Specifically, we apply a simple but analytical toy model, previously discussed in \cite{Baumgarte:2006ug} and motivated by \cite{Pfeiffer:2005jf}, to a cosmological context in order to demonstrate surprising consequences of the non-linearities in the Hamiltonian constraint.  In particular, we show that solutions to the Hamiltonian constraint exist only for certain combinations of the deviations of the mass-energy density and extrinsic curvature from the asymptotic cosmology.  For example, assuming that the extrinsic curvature remains unchanged from the cosmological background will allow for density inhomogeneities only up to a surprisingly small maximum value for horizon scale modes.  Stated differently, this implies that the intrinsic curvature alone cannot balance arbitrarily large inhomogeneities in the mass-energy density.  Moreover, we demonstrate that, if solutions exist, they are not unique.  Instead, there exist two branches of solutions for allowed combinations of the deviations in the mass-energy density and extrinsic curvature.  One of these branches corresponds to solutions with weak intrinsic curvature and approaches the FLRW spacetime in the limit of small inhomogeneities, while the second branch corresponds to strong intrinsic curvature and may feature black-hole horizons and throat-like structures in the initial data.  While both branches represent viable initial data in the sense that they satisfy the Hamiltonian constraint, they may or may not not be astrophysically realistic, in the sense of describing a time-like slice of the universe as it may have emerged from its prior evolution. 

This paper is organized as follows.  In Sect.~\ref{sec-calcs} we adopt the toy model of \cite{Baumgarte:2006ug} in a cosmological setting and provide analytical solutions for two families of solutions.  In Sect.~\ref{sec-diag} we examine the physical properties of these solutions, and in Sect.~\ref{sec-discussion} we briefly discuss potential implications of the results.   Throughout this paper we adopt geometrized units with $c = G = 1$, and the ``standard" 3+1 formalism commonly used in numerical relativity applications (see, e.g., \cite{Baumgarte:2010ndz}).

\section{A toy model for the Hamiltonian constraint}
\label{sec-calcs}

A perfect FLRW spacetime allows a preferred coordinate system in which the metric and stress-energy tensor depend on time only.  The proper distance between these preferred coordinates can be parametrized by a scale factor $a(t)$.  The (relative) expansion rate $H \equiv \dot{a}/a$ then satisfies Friedmann's equation
\begin{equation}
\left( \frac{\dot a}{a} \right)^2 = \frac{8 \pi}{3} \rho_0,
\end{equation}
where $\rho_0 \equiv u_a u_b T^{ab}$ is the mass-energy density as observed by the coordinate observers, $u^a$ their four-velocity, and $T_{ab}$ the stress-energy tensor.  Without loss of generality we may choose $a = 1$ at some initial time $t_0$.

In the context of a 3+1 decomposition we identify hypersurfaces of constant coordinate time $t$ with a spatial slice.  We denote the normal vector on these slices with $n^a$ and the induced spatial metric with $\gamma_{ab} = g_{ab} + n_a n_b$, where $g_{ab}$ is the spacetime metric.  In the preferred coordinate system of an FLRW spacetime discussed above, the normal vectors $n^a$ coincide with the four-velocities $u^a$ of coordinate observers (see, e.g., Exercise 4.1 in \cite{Baumgarte:2010ndz}).

In the following we will focus on spherically symmetric spaces, which are always conformally flat, so that we may write $\gamma_{ij} = \psi^4 \eta_{ij}$.  Here $\psi$ is the conformal factor and $\eta_{ij}$ the flat metric. While this choice simplifies our analysis, we expect our findings to hold qualitatively even in the absence of spherical symmetry or conformal flatness (see, e.g., \cite{Pfeiffer:2005jf} for a numerical example, albeit for asymptotically flat spacetimes, that assumes neither spherical symmetry nor conformal flatness).  For the preferred slicing of a FLRW spacetime we may then identify $\psi = a^{1/2}$ (with $\psi = 1$ on the initial slice at time $t_0$); moreover, the mean curvature $K$, i.e.~the trace of the extrinsic curvature $K = \gamma^{ij} K_{ij}$, which is related to the time-derivative of the spatial metric, is given by 
\begin{equation} \label{eq:friedconst}
K_0 = - 3 H = - 3 \dot{a} / a = -(24 \pi \rho_0)^{1/2}.
\end{equation}

We now set up a scenario in which a local, spherically symmetric over-dense region is embedded at the center of an otherwise homogeneous cosmological spacetime.  Specifically, we consider an initial (spatial) slice at coordinate time $t = 0$, say, also assumed to be spherically symmetric.  At large distances from the center we choose the slice to approach that of an FLRW spacetime, i.e.~with $K \rightarrow -(24 \pi \rho_0)^{1/2}$ and $\psi \rightarrow 1$.  Following \cite{Baumgarte:2006ug} we explore properties of solutions to the Hamiltonian constraint by adopting a toy model with a constant over-density inside some (coordinate) radius $\rod$; i.e.~we assume that the mass-energy density $\rho = n_a n_b T^{ab}$
is given by
\begin{equation}
    \rho(r) = \left\{ \begin{array}{ll}
    \rho_0 + \Delta \rho ~~~~ &  {\rm for} ~r < \rod \\
    \rho_0 \quad & {\rm for} ~r > \rod.
    \end{array}
    \right.
\end{equation}
Here $\rho_0$ is the density of the corresponding asymptotic FLRW spacetime, and $\Delta \rho$ the over-density.  
We make no assumptions on the size of $r_{od}$; the over-dense region may or may not be fully inside the Hubble volume, $\sim H^{-3}$, of the associated asymptotic spacetime.  
The Hamiltonian constraint then becomes
\begin{equation} \label{ham1}
\bar \nabla^2 \psi - \frac{1}{12} \psi^5 K^2 + \frac{1}{8} \psi^{-7} \bar A_{ij} \bar A^{ij} = - 2 \pi \psi^5 \rho,
\end{equation}
where $\bar \nabla^2$ is the flat-space Laplace operator and 
\begin{equation}
 \bar A_{ij} = \psi^2 A_{ij} = \psi^2 \left(K_{ij} - \frac{1}{3} \gamma_{ij} K \right)
\end{equation}
the conformally related trace-free part of the extrinsic curvature  

Before proceeding we note the ``wrong-sign" issue in the Hamiltonian constraint \eqref{ham1}: the fact that the source term $-2 \pi \psi^5 \rho$ on the right-hand side has an overall negative sign with positive exponent of $\psi$ prevents the application of the maximum principle to prove the existence or uniqueness of solutions (see, e.g., \cite{Yor79}).  In \cite{Baumgarte:2006ug}, the absence of solutions for densities larger than a critical density -- together with the non-uniqueness of solutions for smaller densities -- was demonstrated with a simple analytical toy model.  Here we apply the same toy model in a cosmological context, and demonstrate that solutions exist only if departures from homogeneity satisfy certain conditions.

In many numerical relativity applications the wrong-sign issue is avoided by introducing a conformal rescaling $\bar \rho = \psi^n \rho$ of the density, with $n \geq 5$, designed to flip the sign of the exponent of $\psi$ on the right-hand side of the Hamiltonian constraint \eqref{ham1}, and specifying $\bar \rho$ rather than $\rho$.  In that case, a solution for $\psi$ exists for any value of $\bar \rho$. However, if one reconstructs $\rho$ from $\bar{\rho}$ one finds that the limits on the size of $\rho$ have not changed from those presented in \cite{Baumgarte:2006ug}.  This effect is most important when the size of the over-dense region is large.  Here we choose to study solutions of the Hamiltonian constraint for given values of the physical density, rather than the rescaled density, since that is the quantity that can be directly computed from the contents of the Universe.  

We will assume that the mean curvature is still given by the FRLW value \eqref{eq:friedconst} in the exterior of the over-dense region.  In the interior, however, we are free to choose a value for $K$ different from $K_0$; specifically, we will choose 
\begin{equation} \label{eq:K}
    K(r) = \left\{ \begin{array}{ll}
    K_0 + \Delta K ~~~~ &  {\rm for} ~r \leq \rod \\
    K_0 \quad & {\rm for} ~r > \rod.
    \end{array}
    \right.
\end{equation}
Choosing $\Delta K \ne 0$ results in a discontinuity in $K$ at $r = \rod$, meaning that the momentum constraint 
\begin{equation} \label{mom}
    \bar{D}_j \bar A^{ij} =\frac{2}{3} \psi^6 \bar \gamma^{ij} \bar D_j K + 8\pi \psi^{10}  S^i 
\end{equation}
is no longer satisfied analytically with $\bar A_{ij} = 0$ and a vanishing momentum density, $S^i = 0$.  In the following we assume that the contributions from the derivative of $K$ are fully balanced by an appropriate choice of $S^r$, treating it as a fluid for which the momentum density and the energy density are independent degrees of freedom.\footnote{Using a fluid treatment simplifies the process of simultaneously solving the Hamiltonian and momentum constraints; in instances where the energy density $\rho$ and the momentum density $S_i$ are both functions of a common object, say a scalar field, one has to solve these constraints self-consistently.  The physical density $\rho$ associated with a scalar field will still satisfy the Hamiltonian constraint, however, and therefore will still be subject to the ``wrong-sign" issue discussed above.  Accordingly, there will be restrictions on such initial data similar to those discussed here even in that case.}  We may therefore continue to choose $\bar{A}_{ij}=0$, and only concern ourselves with the Hamiltonian constraint.  Allowing a non-vanishing $\bar A_{ij}$ would affect our solutions quantitatively, but, because the ``wrong-sign" issue discussed above persists for non-zero $\bar A_{ij}$, we expect our conclusions to hold qualitatively even in that case (see also \cite{Pfeiffer:2005jf} for a numerical example).

We would like to emphasize that we do not intend to promote this simple toy model -- with an infinite momentum density on an infinitely thin shell -- as a realistic cosmological model.  Instead, this toy model allows us to study properties of solutions to the Hamiltonian constraint in a context in which it can be solved analytically.    The existence of these analytic solutions allow us to characterize the solutions of the Hamiltonian constraint; these conclusions should then be applicable to other more realistic scenarios for which spherical over-dense initial conditions are employed.

It is convenient to define dimensionless quantities to parametrize the over-dense region
\begin{equation}
\delta \rho \equiv \Delta \rho / \rho_0,~~~~~
\delta K \equiv \Delta K / K_0.
\end{equation}
We make no assumptions about the size of these parameters since we will not employ them as order parameters in any expansion. Using these, the Hamiltonian constraint takes the form 
\begin{equation} \label{ham2}
    \bar \nabla^2 \psi + 2 \pi \rho_0 \left( \delta \rho - \delta K (2 + \delta K)\right) \psi^5 =0.
\end{equation}
This equation is the analog of Eq.~(7a) of \cite{Baumgarte:2006ug}.  Here, the non-linear equation for the conformal factor is sourced by an {\em effective} energy density, $\delta \rho - \delta K (2 + \delta K),$ which is a combination of $\delta \rho$ and $\delta K$, rather than just $\rho_0$ (as it is in \cite{Baumgarte:2006ug}).  While Eq.~\eqref{eq:friedconst} allows for $\delta K$ to have either sign, a cosmologically motivated choice would be to take $\delta K$ and $K_0$ to both be negative, to preserve the intuition that larger $\rho$ leads to a larger expansion rate on scales comparable to the horizon.  Note that, by choosing $\delta K$ to have the same sign as $K_0$, the contribution from the cosmological expansion or contraction always decreases the effective energy density that acts as source for the conformal factor.

For a given $\delta \rho$, together with a choice for $\delta K$, we can solve the Hamiltonian constraint (\ref{ham2}) for the conformal factor $\psi(r)$ subject to the boundary condition $\psi(r) \rightarrow 1$ as $r\rightarrow \infty$.  Regular solutions (that remain finite at the origin $r = 0$) can be constructed from two separate analytical solutions for the interior and exterior of the over-dense region, namely
\begin{equation} \label{eq:psi}
    \psi(r) = \left\{ \begin{array}{ll}
    C u_\alpha(r) ~~~~ &  {\rm for} ~r \leq \rod \\
    \beta/r + 1 & {\rm for} ~r > \rod,
    \end{array}
    \right.
\end{equation}
where $C$, $\alpha$, and $\beta$ are constants (not to be confused with the lapse function and the shift vector) and where the Sobolev functions
\begin{equation} \label{eq:sobolev}
    u_\alpha (r) = \sqrt{\frac{\alpha \rod}{r^2 + (\alpha \rod)^2}}
\end{equation}
satisfy the equation
\begin{equation} \label{sobolev}
    \bar \nabla^2 u + 3 u^5 =0
\end{equation}
(see \cite{Baumgarte:2006ug}).  Matching (\ref{sobolev}) with (\ref{ham2}) we see that the constant $C$ in (\ref{eq:psi}) is given by
\begin{equation} \label{eq:C}
    C = \left[\frac{2\pi \rho_0}{3} \left( \delta \rho - \delta K(2 + \delta K) \right)\right]^{-1/4}.  
\end{equation}

\begin{figure}
\centering
\includegraphics[width=0.45\textwidth]{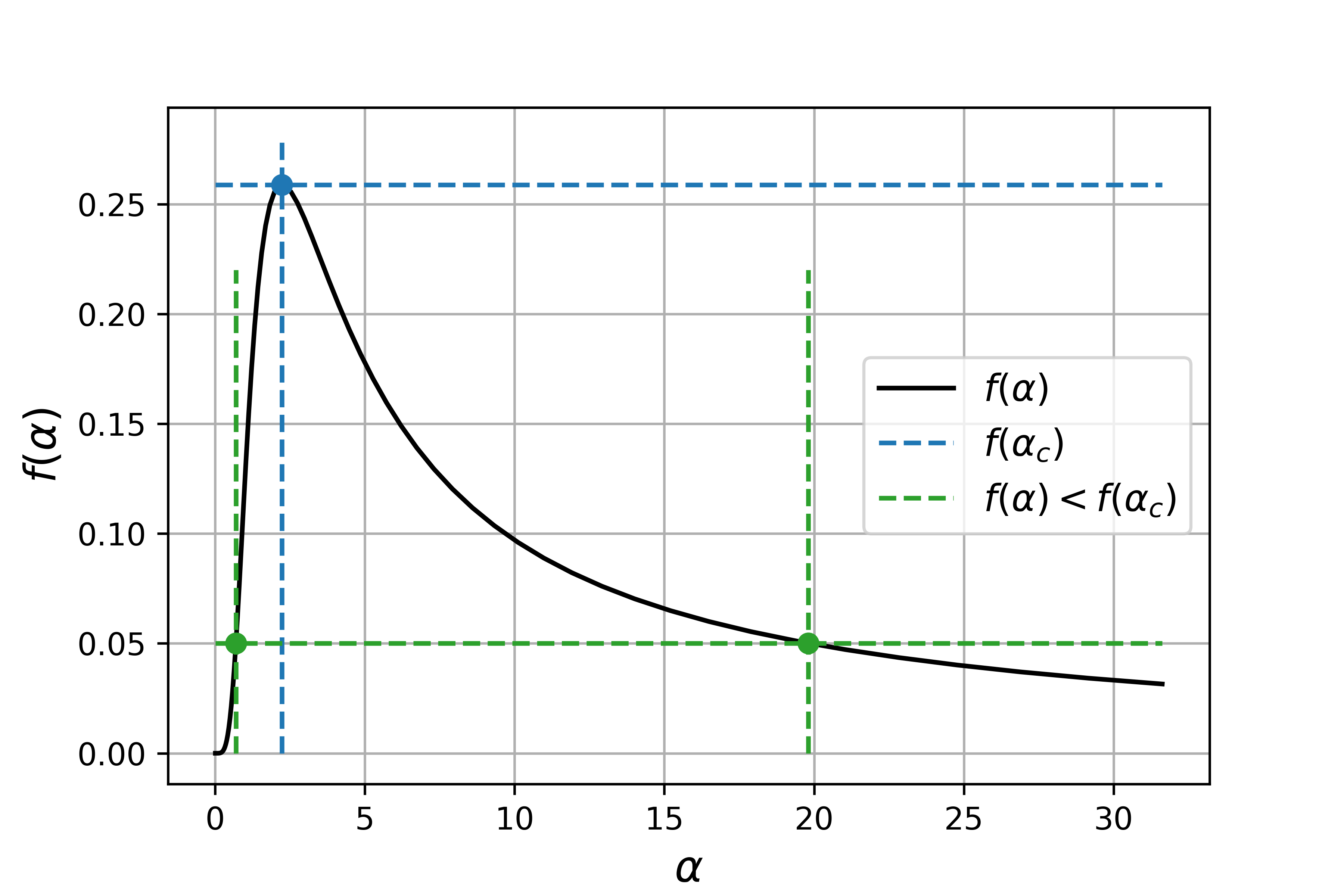}
\caption{The function $f(\alpha)$ defined in Eq.~(\ref{eq:f_alpha}). There are no solutions with $f(\alpha)$ greater than the value at the maximum of the function $f(\alpha_c)$. For smaller values of $f(\alpha)$ there are two solutions for $\alpha$, identified with the strong, $\alpha < \alpha_c$, and weak, $\alpha > \alpha_c$, branches of the solution.}
\label{fig-falpha}
\end{figure}

We next determine the constants $\alpha$ and $\beta$ by requiring that both $\psi(r)$ and its first derivative be continuous at $r = \rod$.  Continuity of $\psi$ results in the condition 
\begin{equation} \label{eq:beta}
    \beta = R(Cu_\alpha(\rod) - 1)~,
\end{equation}
while continuity of the derivative yields 
\begin{align} 
    f(\alpha) & \equiv \frac{\alpha^5}{(1+ \alpha^2)^3} = \frac{\rod}{C^2}
    \nonumber \\
    & = \left( \frac{2 \pi \rho_0 \rod^2}{3} \right)^{1/2} \left(  \delta \rho - \delta K(2 + \delta K) \right)^{1/2}.
    \label{eq:f_alpha}
\end{align}
Inserting this into \eqref{eq:beta} we see that $\beta$ takes the simple form
\begin{equation} \label{eq:beta2}
\beta = \frac{\rod}{\alpha^2}.
\end{equation}
We show a graph of the function $f(\alpha)$ in Fig.~\ref{fig-falpha}.  Note, in particular, that the function tends to zero for both $\alpha \rightarrow 0$ and $\alpha \rightarrow \infty$, and that it takes a maximum of $f(\alpha_c) = 5^{5/2} / 6^3$ at the critical value $\alpha_c =\sqrt{5}$.  For a given value of $\rod$, solutions to the Hamiltonian constraint therefore exist only when 
\begin{equation} \label{lim1}
\delta \rho - \delta K(2 + \delta K) \leq \frac{5^5}{6^6} \frac{3}{2 \pi \rho_0 \rod^2}.
\end{equation}

In order to further examine the above limits it is useful specify the size of the over-dense region in terms of the Hubble scale $L_H = 3/|K_0|$
associated with the asymptotic spacetime, i.e.~we write\footnote{We note that $n$ need not be an integer.} 
\begin{equation} \label{eq:R_hubble}
\rod = n L_H = \frac{3 n}{|K_0|} = \frac{3 n}{\sqrt{24 \pi \rho_0}} = \frac{1}{2} \left( \frac{3}{2 \pi \rho_0} \right)^{1/2} n  
\end{equation}
so that 
\begin{equation}
    \left(\frac{2 \pi \rho_0 \rod^2}{3}\right)^{1/2} = \frac{n}{2}. 
\end{equation}
Inserting this into (\ref{lim1}) we obtain
\begin{equation} \label{eq:rhocrit}
\delta \rho \leq \delta \rho_{\rm crit} = \frac{5^5}{6^6} \frac{4}{n^2} + \delta K(2 + \delta K).
\end{equation}
Evidently, solutions to the Hamiltonian constraint exist only for $\delta \rho$ up to a certain maximum value that depends sensitively on the choice of the mean curvature inside the over-dense region.

We may interpret the different terms in (\ref{eq:rhocrit}) from their origin in the Hamiltonian constraint (\ref{ham1}).  In the latter, any density source $\rho$ on its right-hand side must be balanced by the combination of the {\em intrinsic curvature} term $\bar \nabla^2 \psi$ and the {\em extrinsic curvature} term $\psi^5 K^2 / 12$ on the left-hand side.  In \eqref{eq:rhocrit}, the first term on the right-hand side originates from the intrinsic curvature, and the second from the extrinsic curvature.  Naively, one might have expected that any given $\delta \rho$ can be balanced by either intrinsic or extrinsic curvature, or some arbitrary combination of both.  The constraint \eqref{eq:rhocrit} demonstrates, however, that this is {\em not} the case.  Instead, the intrinsic curvature can only accommodate some choices for the size of the over-density, and larger excursions  have to be balanced with an inhomogeneous extrinsic curvature.  

Specifically, if the over-dense region is of Hubble scale, $n \simeq 1$, the maximum value of $\delta \rho$  that can be accommodated by the intrinsic curvature alone, i.e.~with $\delta K = 0$, is given by
\begin{equation}
    \delta \rho_{\rm crit} \simeq 0.27 \quad (n = 1,~\delta K = 0).
\end{equation}
Larger values of $\delta \rho$ can be constructed only by allowing the local expansion (or contraction) to increase accordingly.  The constraint \eqref{eq:rhocrit} is even more restrictive for super-horizon scales,  $n > 1$.  In this case, the over-dense region must either satisfy $\delta \rho \lesssim 0.27/n^2$, or the extrinsic curvature must play a significant role in order for solutions to the Hamiltonian constraint \eqref{ham1} to exist.

\begin{figure}
\centering
\includegraphics[width=0.45\textwidth]{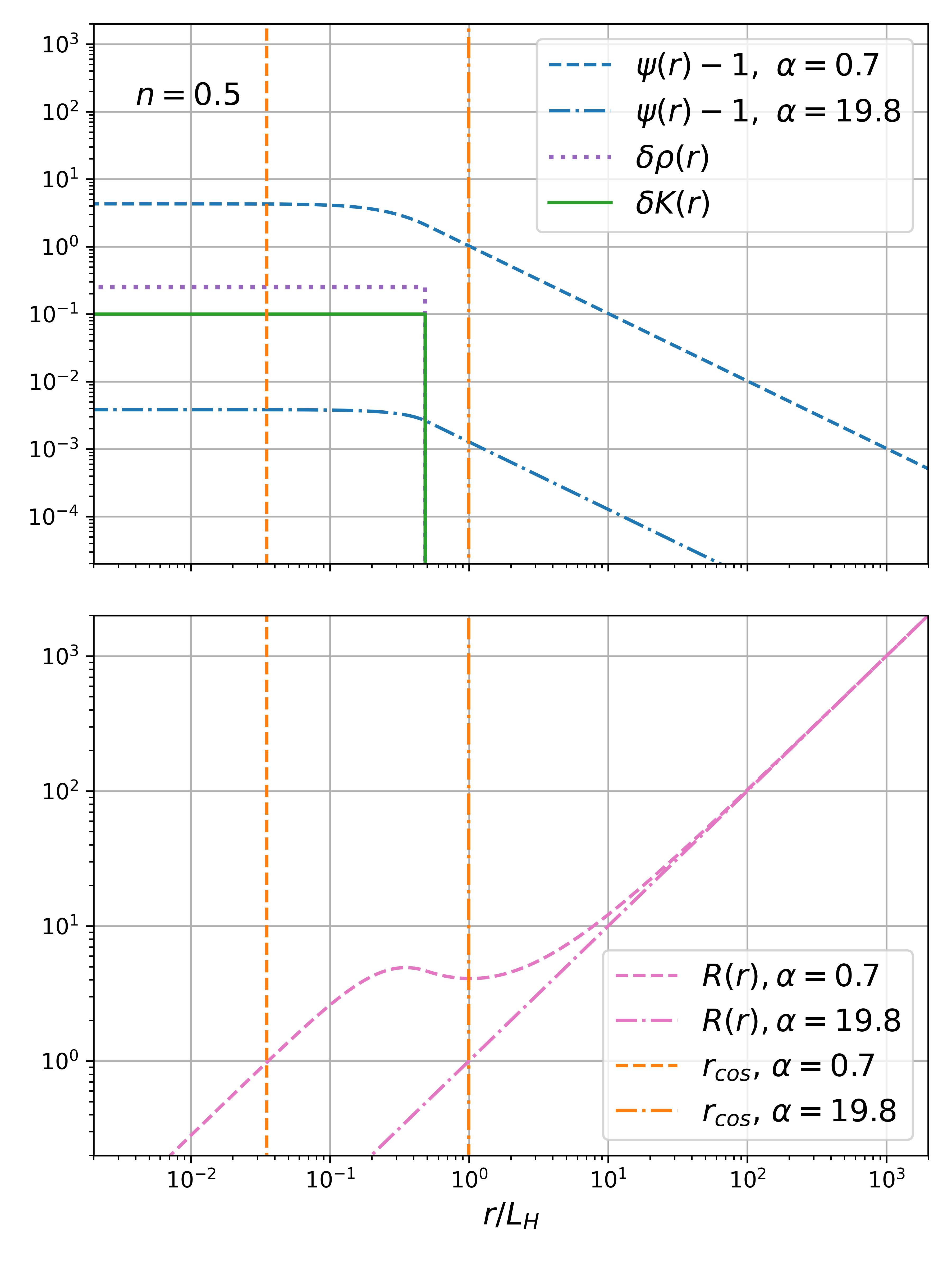}
\caption{
Top panel: Representative solutions $\psi(r)$ for a fixed value of the additional expansion $\delta K = 0.1$, a Hubble size region $n=0.5$ and the two values of $\alpha$ from the strong and weak branches as marked on Fig.~\ref{fig-falpha}, namely $\alpha=0.7$ and $\alpha = 19.8$. Both values give $f(\alpha)=0.05$ and so the same profile for $\rho$, but the profiles for $\psi$ differ.  The vertical lines show the coordinate locations of the cosmological horizons for the two solutions.
Bottom panel: The areal radius $R = \psi^2 r$ as a function of the coordinate radius $r$ for the two solutions.  Although the coordinate radii of the cosmological horizons are different in each solution, both horizons have the same areal radius $R(\rcos)=1$ and therefore the same entropy. They are, however, physically distinct, with those on the strong branch exhibiting a non-monotonic behavior in the areal radius $R$ as a function of coordinate radius $r$ in cases with $\alpha<1$.}
\label{fig-IC}
\end{figure}

To recap, we specify a (constant) spherical over-density $\delta \rho$ of size $\rod = n L_H$ together with a value for the (constant) mean-curvature deviation  $\delta K$.  For these choices, solutions to the Hamiltonian constraint \eqref{ham1} exist only if $\delta \rho < \delta \rho_{\rm crit}$ as defined in Eq.~\eqref{eq:rhocrit}, so that the right-hand side of Eq.~\eqref{eq:f_alpha} is less than the maximum value of the function $f(\alpha)$.  Note, however, that if these solutions exist, they are not unique.  Specifically, we can construct two different solutions based on the two values of $\alpha$ corresponding to any (positive) value of $f(\alpha) < f(\alpha_c)$, as illustrated in Fig.~\ref{fig-falpha}.  For each of the two allowed values of $\alpha$ we can construct the solution for $\psi(r)$ from \eqref{eq:psi}.  Accordingly, we find two different branches of solutions, as illustrated in Fig.~\ref{fig-IC}.   We analyze the properties of the two branches of solutions in the following section. 

\section{Physical characteristics of the solutions}
\label{sec-diag}

There are two relevant diagnostics that we can use to characterize the solutions found in Sect.~\ref{sec-calcs}, namely the Misner-Sharp mass and the location (or absence of) apparent horizons.

\subsection{Misner-Sharp Mass}
\label{sec-mass}

The Misner-Sharp mass function $M_{\rm MS}(r)$ reduces to the Schwarzschild mass for Schwarzschild spacetimes and to the enclosed mass $M_{\rm MS}(r) = 4\pi (ar)^3 \rho/3$ at coordinate radius $r$ for FLRW spacetimes with scale factor $a$.    In the spherically symmetric metric described above, with $\gamma_{ij} = \psi^4 \eta_{ij}$, the Misner-Sharp mass \cite{Misner:1964je} is given by
\begin{equation} \label{MS_mass}
    M_{\rm MS}(r) =  \frac{r^3 \psi^6 K^2}{18} - 2r^3(\partial_r \psi)^2 - 2r^2 \psi \partial_r \psi
\end{equation}
(we follow \cite{Thornburg:1998cx} and use the formulation in Misner, Thorne and Wheeler \cite{1973grav.book.....M}, Sect.~23.5.)

\begin{figure}
    \centering    \includegraphics[width=0.45\textwidth]{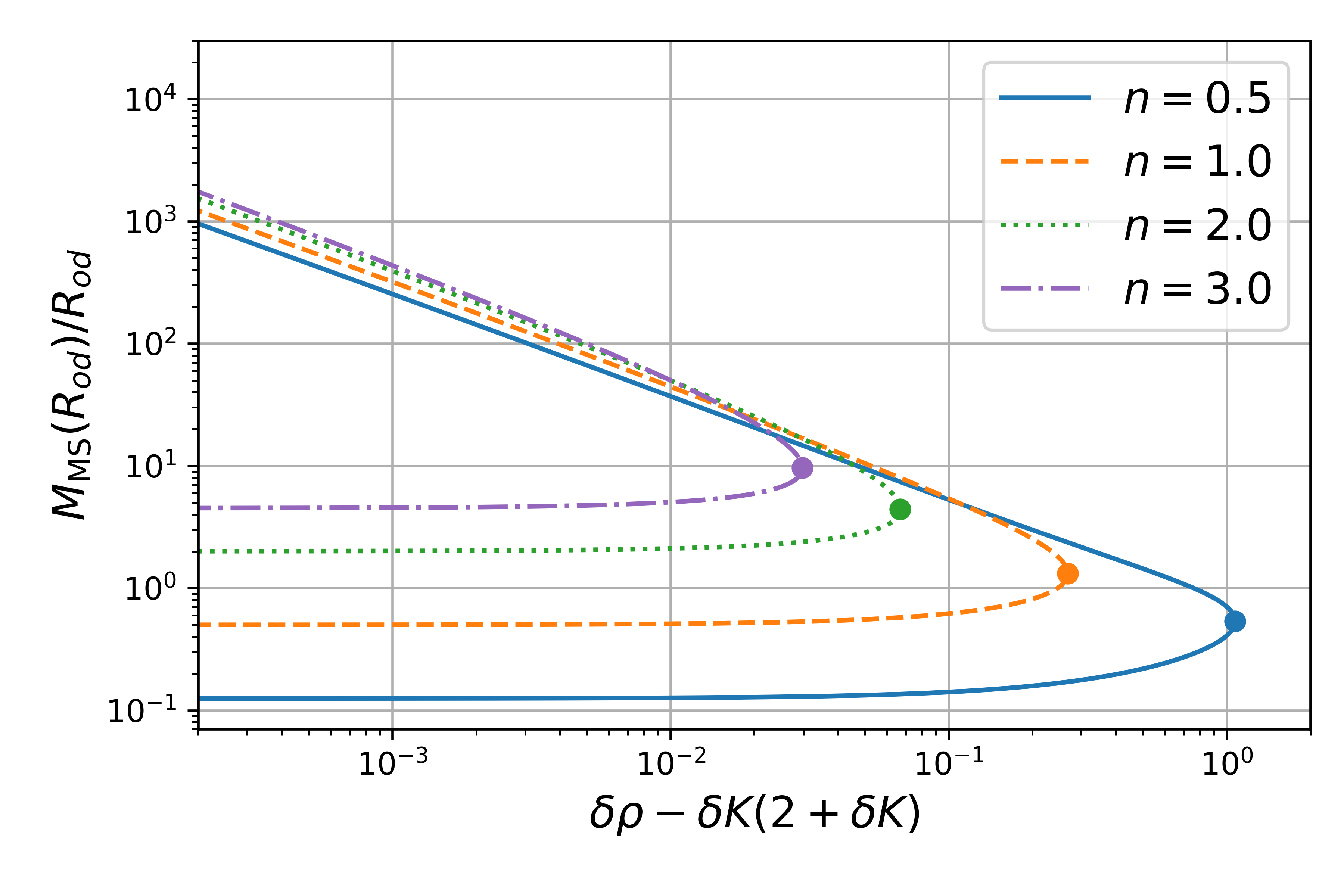}
    \caption{
    Values of the compaction \eqref{eq:compaction} versus $\delta \rho - \delta K(2 + \delta K)$ for different values of $n$.  For a given value of $n$, solutions to the Hamiltonian constraint \eqref{ham1} exist only for suitable combinations of the inhomogeneities $\delta \rho$ and $\delta K$.  If such a solution exists, it is not unique: instead, the Hamiltonian constraint allows both a weak-field solution (the lower branch) and a strong-field solution (the upper branch).  The two branches meet at the critical solution (corresponding to $\alpha = \alpha_{\rm crit}$), marked by the filled dots.      }
    \label{fig:mass}
\end{figure}

In the exterior of the over-dense region, in particular, the Misner-Sharp mass \eqref{MS_mass} becomes
\begin{equation}
M_{\rm MS}(r) = 2 \beta + \frac{4 \pi \rho_0}{3} \frac{(r + \beta)^6}{r^3},
\end{equation}
which reduces to the expected results in both of the limits $\beta \rightarrow 0$ (an FLRW spacetime) and $\rho_0 \rightarrow 0$ (a Schwarzschild spacetime).  Evaluating this expression at the surface $\rod$ of the over-dense region, and inserting \eqref{eq:beta2} and \eqref{eq:R_hubble}, we obtain 
\begin{align} \label{eq:compaction}
    \frac{M_{\rm MS}(\rod)}{\rodprop} & = 
    \frac{M_{\rm MS}(\rod)}{\psi^2 \rod} \nonumber \\
    & = \frac{2 \alpha^2}{(\alpha^2 + 1)^2} + \frac{n^2}{2} \left(1 + \frac{1}{\alpha^2} \right)^4
\end{align}
for the compaction of the over-dense region,  where $\rodprop = \psi^2 \rod$ is the areal radius of the over-dense region. Note that $M_{\rm MS}/\rodprop \rightarrow \infty$ in the limit $\alpha \rightarrow 0$, indicating that small values of $\alpha$ correspond to strong-field solutions, while large values of $\alpha$ correspond to weak-field solutions.  In Fig.~\ref{fig:mass} we show values of the compaction \eqref{eq:compaction} versus the effective energy density, $\delta \rho - \delta K(2 + \delta K)$ (from \eqref{eq:f_alpha}) for families with different values of $n$.   

In the bottom panel of Fig.~\ref{fig-IC} we also show the areal radius $\rprop = \psi^2 r$ as a function of the coordinate radius $r$ for the two representative examples of the strong-field and weak-field solutions high-lighted in Fig.~\ref{fig-falpha}.  In particular, notice that the areal radius monotonically increases with $r$ for the weak-field solution, but features a non-monotonic behavior for the strong-field solution.  In \cite{Uehara:2025idq}, this behavior was used to identify ``Type-II" initial data, which suggests that the initial data used in that work belong to what we refer to as strong-field branch data.

\subsection{Horizons}
\label{sec-horizon}

In order to identify horizons we compute the expansion of both outgoing and ingoing null geodesics, which yields\footnote{Note that the subscripts ``out" and ``in" refer to the direction of null geodesics, and not to the interior or exterior of the over-dense region.}
\begin{subequations} \label{eq:exp}
\begin{align} 
    \Theta_{(\rm out)}(r) & = \frac{2}{r \psi^2} + \frac{4 \partial_r \psi}{\psi^3} - \frac{2K}{3}, \label{eq:exp_out} \\
    \Theta_{(\rm{in})}(r) & = - \frac{2}{r \psi^2} - \frac{4 \partial_r \psi}{\psi^3} - \frac{2K}{3},
    \label{eq:exp_in}
\end{align}
\end{subequations}
respectively (see, e.g., \cite{1989fnr..book...89Y}).  
We then identify trapped surfaces with regions where inward and outward null expansions are negative. The black-hole apparent horizon $r_{\rm BH, out}$ is the marginally outer trapped surface, the outermost surface where the outgoing null expansion is zero, i.e.~the largest value of $r$ for which $\Theta_{(\rm{out})} = 0$.  We similarly identify a cosmological apparent horizon $\rcos$ with surfaces for which the ingoing null expansion $\Theta_{(\rm{in})} = 0$.
(See also \cite{Faraoni:2015ula} for a discussion of different types of horizon).

We will also evaluate the entropy of the cosmological horizons \cite{Gibbons:1977mu}.  Whilst this is a dynamical case, not a stationary one, and so the entropy is not well-defined, in spherical symmetry recent arguments have suggested that it may be approximated by the proper area of the apparent horizon (see most recently \cite{Wald:2025nbz}, and key works on dynamical horizons \cite{Ashtekar:2003hk,Ashtekar:2005ez,Bousso:2015mqa})
\begin{equation}
    S_H =4 \pi R(\rcos)^2 ~.
\end{equation}
We find that this measure of entropy of cosmological horizons in the interior of the over-dense region takes the same value on both branches of solutions, even though the solutions are otherwise physically different (see Eq.~\eqref{eq:cos_hor_proper} below).

Before proceeding we evaluate the first two terms on the right-hand sides of Eqs.~(\ref{eq:exp}) in both the interior and the exterior of the over-dense region from the respective solutions (\ref{eq:psi}), which yields
\begin{equation} \label{eq:first_terms}
    \frac{2}{r \psi^2} + \frac{4 \partial_r \psi}{\psi^3} = 
    \left\{ \begin{array}{ll}
    \displaystyle
    \frac{2}{\alpha \rod C^2} \left(\frac{\alpha^2 \rod^2}{r} - r \right) ~~~~ & \mbox{for}~r \leq \rod, \\
    \displaystyle \frac{2 r}{(\beta + r)^3} \left(r - \beta \right)
    & \mbox{for}~r > \rod. 
    \end{array}
    \right.
\end{equation}
We will similarly use 
\begin{equation} \label{eq:K2}
    K = 
    \left\{ \begin{array}{ll}
    \displaystyle
    K_0( 1 + \delta K) ~~~~ & \mbox{for}~r \leq \rod, \\
    K_0
    & \mbox{for}~r > \rod, 
    \end{array}
    \right.
\end{equation}
and will assume an expanding Universe with 
\begin{equation}
    K_0 < 0
\end{equation}
in the following, unless noted otherwise.

\subsubsection{Black hole horizons}

Inserting \eqref{eq:first_terms} into (\ref{eq:exp_out}) we find
\begin{equation} \label{eq:theta_hor}
    \Theta_{(\rm out)} = 
    \left\{ \begin{array}{ll}
    \displaystyle
     \frac{2}{\alpha \rod C^2} \left(\frac{\alpha^2 \rod^2}{r} - r \right) - \frac{2 K}{3}~~~~ & \mbox{for}~r \leq \rod, \\
    \displaystyle \frac{2 r}{(\beta + r)^3} \left(r - \beta \right) - \frac{2 K}{3}
    & \mbox{for}~r > \rod, 
    \end{array}
    \right.
\end{equation}
We first observe that $\Theta_{(\rm out)} > 0$ for both $r \rightarrow 0$ and $r \rightarrow \infty$, meaning that the expansion of outgoing null geodesics can only have an {\em even} number of roots $\Theta_{(\rm out)} = 0$.  In particular, the immediate vicinity of the origin will never be trapped; if there is a trapped region, it will extend between two non-zero values of the radius $r$.   

In the interior of the over-dense region, the condition $\Theta_{(\rm out)} = 0$ reduces to the quadratic equation
\begin{equation} \label{eq:ah_hor_int}
    r^2 + \frac{1}{3} K C^2 \alpha \rod r - \alpha^2 \rod^2 = 0.
\end{equation}
The solution are then given by 
\begin{equation} \label{eq:hor_int1}
    r_{\rm BH, int} = \alpha \rod
    \left(A \pm \left( 1 + A^2 \right)^{1/2} \right),
\end{equation}
where we have defined the dimensionless positive constant
\begin{align} \label{eq:A}
    A & \equiv -\frac{KC^2}{6} = - \frac{K \rod}{6 f(\alpha)} = \frac{n}{2 f(\alpha)} ( 1 + \delta K) \nonumber \\
    & = \frac{n}{2} \frac{(1+\alpha^2)^3}{\alpha^5} (1 + \delta K.)
\end{align}
Evidently, only the positive root in \eqref{eq:hor_int1} can yield a positive radius, so that $\Theta_{(\rm out)}$ can have only one root in the interior of the over-dense region.  Since this root cannot be the only root of $\Theta_{(\rm out)}$, as we argued above, we see that, if a black hole apparent horizon exists, it must be in the exterior of the over-dense region.  

We also note that \eqref{eq:hor_int1} provides a valid solution only if $r_{\rm BH} \leq \rod$, which is equivalent to the condition
\begin{equation} \label{eq:A_cond}
    \alpha \left( A + (1 + A^2)^{1/2} \right) < 1. 
\end{equation}
For $K_0 = 0$, which corresponds to $A = 0$, in particular, black hole horizons exist in the interior of the over-dense region only for the part of the strong-field branch with $\alpha \leq 1$ (see \cite{Baumgarte:2006ug}).  For $K_0 < 0$, however, we see that $A \rightarrow \infty$ in both limits $\alpha \rightarrow 0$ and $\alpha \rightarrow \infty$, so that black-hole horizons can exist in the interior only for intermediate values of $\alpha$.

In the exterior, roots of $\Theta_{(\rm out)}$ can be found from the cubic equation
\begin{equation} \label{eq:cubic1}
    r^2 - \beta r - \frac{1}{3} K_0 (\beta + r)^3 = 0. 
\end{equation}
For $K_0 = 0$ we now have $r_{\rm BH,out} = \beta = \rod / \alpha^2$ as non-zero solutions.  This solution satisfies $r_{\rm BH,out} \geq \rod$ for $\alpha < 1$ only, consistent with our finding above.  For $K_0 = 0$ we therefore find trapped surfaces for $\alpha < 1$ between $r_{\rm BH,int} = \alpha \rod$ and $r_{\rm BH,out} = \rod / \alpha^2$.  For small $K_0$, with $K_0 \rod \ll 1$, corrections to these solutions can be found perturbatively.  

Exact solutions to the cubic equation \eqref{eq:cubic1} can also be found using, for example, the recipe on page 228 of \cite{Press:2007ipz}.  In particular, this involves computing expressions $\bar Q$ and $\bar R$ from the coefficients of the cubic, which, in our case, reduce to\footnote{We use bars on $\bar Q$ and $\bar R$ in order to avoid confusion with the areal radius $\rprop$.}
\begin{equation} \label{eq:QandR}
    \bar Q = \frac{1 - 3K_0\beta}{K_0^2},~~~~\bar R = \frac{2 - 9K_0 \beta + 6 K_0^2\beta^2}{2 K_0^3}
\end{equation}
(see Eq.~(5.6.10) in \cite{Press:2007ipz}).  If $\bar R^2 - \bar Q^3 < 0$, then the cubic equation has three real roots, and otherwise it has only one real root (which do not have to be in the exterior of the over-dense region, of course).  From \eqref{eq:QandR} we have
\begin{equation} \label{eq:R2-Q3}
    \bar R^2 - \bar Q^3 = \frac{9 \beta^4}{K_0^2} - \frac{3 \beta^2}{4 K_0^4} = \frac{\rod^6}{108 n^4 \alpha^8} (108\, n^2 - \alpha^4),
\end{equation}
indicating that we can have multiple real solutions only if 
\begin{equation} \label{eq:alpha_condition}
    \alpha^2 > 6 \sqrt{3}\, n.
\end{equation}
We saw above that, for $n > 0$ and $\alpha$ sufficiently small, no black hole horizon can exist in the interior of the over-dense region.  If there were a trapped region at all, we would then need {\rm two} roots of $\Theta_{(\rm out)}$ in the exterior.  Since this is possible only for values of $\alpha$ satisfying \eqref{eq:alpha_condition}, we conclude that for $K_0 < 0$ and sufficiently small $\alpha$, the initial slice cannot contain any black hole horizons. 

\begin{figure}
    \centering
    \includegraphics[width=0.45\textwidth]{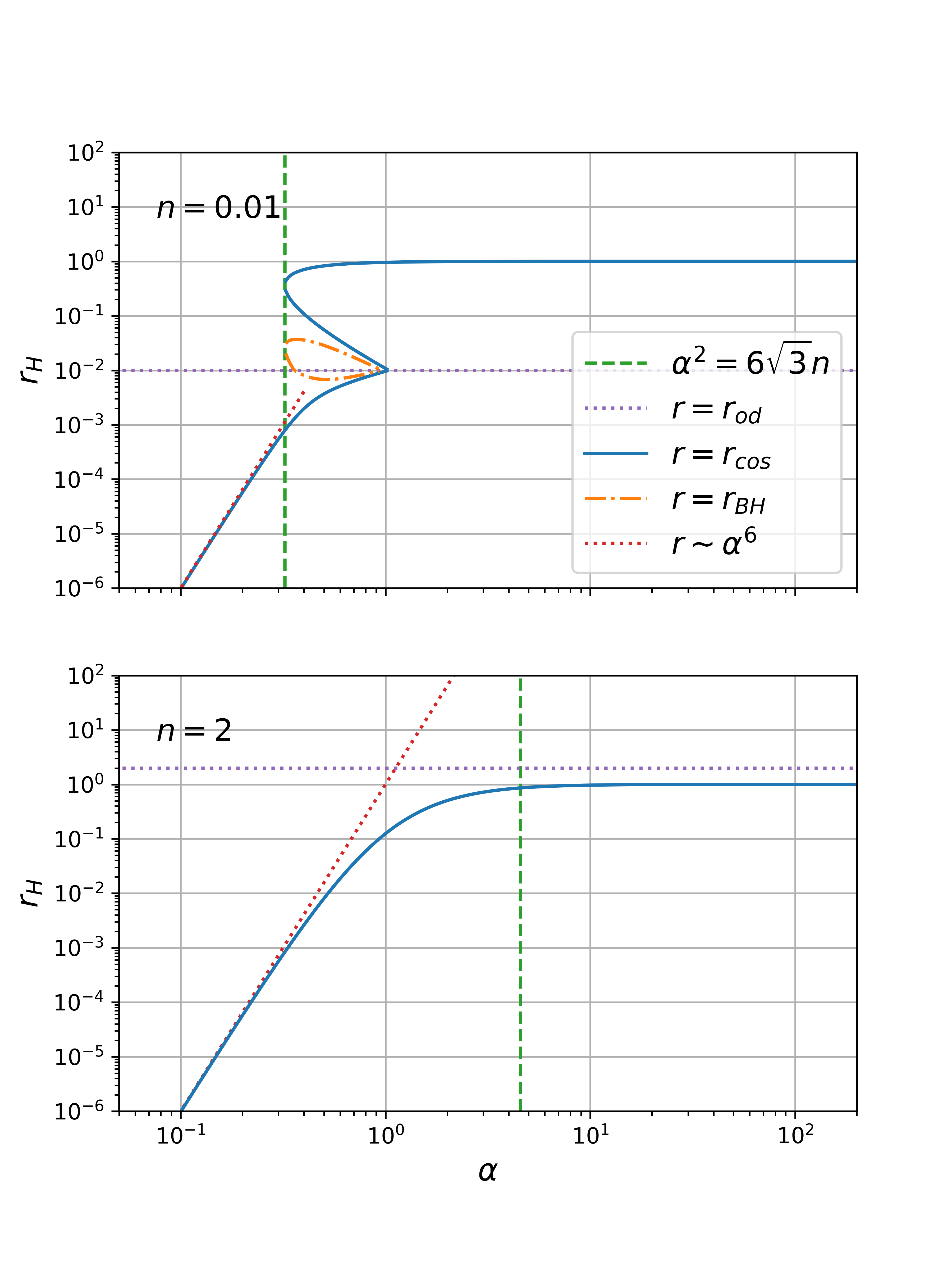}
    \caption{Illustration of the coordinate radius of the horizons for a small $n$ (top panel) and large $n$ (bottom panel) case as a function of the parameter $\alpha$. For a given value of $\alpha$, there is always an even number of black-hole horizons and an odd number of cosmological horizons. In the small $n$ case we have two black hole horizons and three cosmological ones in the region $6\sqrt{3}n < \alpha^2 < 1$ (see Eq.~\eqref{eq:alpha_condition}). For the large $n$ case no BH horizons exist for any value of $\alpha$, and there is only one cosmological horizon. As discussed in the text, whilst the horizons on the strong and weak branches have different coordinate radii, their proper radii are the same for corresponding solutions (see Eq.~\eqref{eq:cos_hor_proper}). The dotted line in the lower panel represents the asymptotic solution \eqref{eq:r_cos_limit}.}
    \label{fig:horizons}
\end{figure}

\subsubsection{Cosmological horizons}

For cosmological horizons we insert Eqs.~\eqref{eq:first_terms} into \eqref{eq:exp_in} to find
\begin{equation} \label{eq:theta_cos}
    \Theta_{(\rm in)} = 
    \left\{ \begin{array}{ll}
    \displaystyle
     - \frac{2}{\alpha \rod C^2} \left(\frac{\alpha^2 \rod^2}{r} - r \right) - \frac{2 K}{3}~~~~ & \mbox{for}~r \leq \rod, \\
    \displaystyle -\frac{2 r}{(\beta + r)^3} \left(r - \beta \right) - \frac{2 K}{3}
    & \mbox{for}~r > \rod, 
    \end{array}
    \right.
\end{equation}
For an expanding universe with $K < 0$ we now have $\Theta_{(\rm in)} < 0$ as $r \rightarrow 0$ but $\Theta_{(\rm in)} > 0$ as $r \rightarrow \infty$, indicating that the initial slice must contain an {\em odd} number of cosmological apparent horizons.

In the interior of the over-dense region, the horizon condition $\Theta_{(\rm in)} = 0$ again reduces to a quadratic equation,
\begin{equation}
   r^2 - \frac{1}{3} K C^2 \alpha \rod r - \alpha^2 \rod^2 = 0
\end{equation}
where $K = K_0(1 + \delta K)$.  In terms of the constant $A$ defined in \eqref{eq:A} the solutions are now given by 
\begin{equation}
\label{eq:hor_int2}
    \rcos = \alpha \rod
    \left(-A \pm \left( 1 + A^2 \right)^{1/2} \right). 
\end{equation}
As for the black hole horizon, we have to pick the positive root for $\rcos$ to be positive.  Unlike for the black hole horizons, however, we find valid solutions with $\rcos < R$ even for large $A$.  Expanding to leading order for $A \gg 1$, \eqref{eq:hor_int1} becomes
\begin{equation}
\label{eq:hor_int_expand}
    \rcos \simeq \frac{\alpha \rod}{2 A}.
\end{equation}

In the strong-field limit $\alpha \rightarrow 0$ we have $A \simeq n (1 + \delta K) / (2 \alpha^5)$, so that \eqref{eq:hor_int_expand} becomes approximately 
\begin{equation}    \label{eq:r_cos_limit}
    \rcos \simeq \alpha^6 \frac{\rod}{n} \frac{1}{1+ \delta K}  ~~~~(\alpha \rightarrow 0).
\end{equation}
Even though this (coordinate) radius approaches zero as $\alpha \rightarrow 0$, the horizon's areal radius does not. To see this, we compute, for any value of $\alpha$,
\begin{align} \label{eq:cos_hor_proper}
    \rcosprop & = \rcos \psi^2 = \rcos C^2 u^2_{\alpha}(\rcos)
        = \rcos C^2 \frac{\alpha \rod}{\rcos^2 + \alpha^2 \rod^2} \nonumber \\
        & = \frac{\rod}{f(\alpha)} \frac{-A + \sqrt{1 + A^2}}{(-A + \sqrt{1 + A^2})^2 + 1}
        = \frac{\rod}{2 f(\alpha)} \frac{1}{\sqrt{1 + A^2}} \nonumber \\
        & = \frac{\rod}{\sqrt{4 f^2(\alpha) + n^2 (1 + \delta K)^2}} 
\end{align}
where we used $C^2 = \rod / f(\alpha)$ as well as the identity
\begin{equation}
- A + \sqrt{1 + A^2} = \frac{1}{A + \sqrt{1 + A^2}}    
\end{equation}
together with Eqs.~\eqref{eq:psi}, \eqref{eq:sobolev}, \eqref{eq:hor_int2}.

Eq.~\eqref{eq:cos_hor_proper} shows several interesting properties of the areal radius of the cosmological horizon, and hence its proper area and entropy.  We first observe that $f(\alpha) \rightarrow 0$ as $\alpha \rightarrow 0$, so that $\rcosprop$ indeed remains finite in this limit, as we stated above.  The same is true in the weak-field limit $\alpha \rightarrow \infty$, for which $f(\alpha) \rightarrow 0 $ also.   We therefore see that 
\begin{equation}
\rcosprop \simeq \frac{\rod}{n(1+\delta K)}
\end{equation}
in both limits $\alpha \rightarrow 0$ and $\alpha \rightarrow \infty$. 

In fact, the cosmological horizon's areal radius \eqref{eq:cos_hor_proper} depends on $\alpha$ only through the function $f(\alpha)$ defined in \eqref{eq:f_alpha}.  As we discussed there, a given value of $f(\alpha)$, corresponding to a given value of the effective energy density $\delta \rho - \delta K ( 2 + \delta K)$, permits two different solutions for $\alpha$, defining the weak-field and strong-field branches of solutions.  However, since \eqref{eq:cos_hor_proper} depends on $f(\alpha)$ only, and not on $\alpha$ individually, we see that the cosmological horizon's areal radius, for given values of $\delta \rho$ and $\delta K$, takes the same value on the weak-field and strong-field branches, as does its associated entropy.

In the exterior of the over-dense region, roots of $\Theta_{(\rm in)}$ can be found from the cubic equation
\begin{equation} \label{eq:cubic2}
    - r^2 + \beta r - \frac{1}{3} K_0 (\beta + r)^3 = 0. 
\end{equation}
As for the black-hole horizons, we can evaluate the sign of the combination $\bar R^2 - \bar Q^3$ to decide whether there exist one or three real roots.  Since only those terms in the cubic equation that involve $K$ enter this criterion (see \eqref{eq:R2-Q3}), and since those terms are the same in $\Theta_{(\rm out)}$ and $\Theta_{(\rm in)}$, we again recover the same criterion \eqref{eq:alpha_condition}.  This finding is confirmed in Fig.~\ref{fig:horizons}, which demonstrates that pairs of black-hole and cosmological horizons appear in the exterior of the over-dense region for the same values of $\alpha$.

\section{Discussion}
\label{sec-discussion}

In this work, we address solutions to the Hamiltonian constraint for cosmological spacetimes.  These solutions are needed as initial data for cosmological simulations in the presence of non-perturbative inhomogeneities.   In order to explore, for instance, whether primordial black holes may form in the early Universe, or whether inflationary scenarios are robust, one might consider a large range of initial density fluctuations.  According to Einstein's equations, these initial conditions will result in spacetime curvature, expressed by the terms on the left-hand side of the Hamiltonian constraint \eqref{ham1}. 

Naively, one might assume that this spacetime curvature could be chosen to be either curvature {\em intrinsic} to the initial slice, expressed by the first term on the left-hand side of \eqref{ham1}, or by the {\em extrinsic} curvature, expressed by the mean curvature $K$ and corresponding to the current expansion (or contraction) of the Universe.   In this paper we apply an analytical toy model, previously used by \cite{Baumgarte:2006ug} to illustrate properties of solutions to the Hamiltonian constraint, in a cosmological context and demonstrate that solutions to \eqref{ham1} exist only if a combination of $\delta \rho$ and $\delta K$ satisfy the restriction \eqref{eq:rhocrit}.  
Moreover, we demonstrate that these solutions are not unique: for suitable perturbations in the density and expansion, there are two branches of solutions describing possible intrinsic curvature perturbations, corresponding to strong-field and weak-field solutions, respectively.  

While the specific restrictions on the density and extrinsic curvature perturbations derived above hold only for our specific toy model with piecewise constant functions for $\delta \rho$ and $\delta K$, qualitatively similar restrictions will hold for other choices of perturbations.  Our results, therefore, have two important consequences for the construction of cosmological initial data.  

First, solutions to the Hamiltonian exist only up to certain maximum values of the over-density, or combinations of the perturbations in the density and extrinsic curvature.  Stated differently, a sufficiently large over-density cannot be balanced by the intrinsic curvature alone; instead, the rate at which space expands or contracts, expressed by the extrinsic curvature, cannot remain homogeneous in the presence of such an over-density.

Second, for allowed combinations of $\delta \rho$ and $\delta K$, we should expect the existence of two possible solutions to the Hamiltonian constraint.  In particular, this means that a numerical, iterative algorithm designed to construct solutions to the Hamiltonian constraint may converge to either one of the two solutions, depending on the initial guess, leading to potentially inconsistent results in the evolution.   While both solutions may represent viable solutions to the Hamiltonian constraint, they may not conform with our astrophysical expectation regarding the properties of the Universe at the particular time that we try to model.

These conclusions imply consequences for understanding how PBH may have formed.  In the standard treatment, PBH are formed by the collapse of over-dense regions as these regions re-enter the cosmic horizon \cite{Carr:1974nx,Carr:1975qj}.  These over-dense regions are a result of rare excursions of the comoving curvature perturbation, $\zeta = \Psi + \delta\rho/H$, whose statistics are frozen-in at long wavelengths during the inflationary epoch \cite{Mukhanov:1990me,Lyth:2004gb,Malik:2008im}.  In linear perturbation theory, this quantity is gauge-invariant \cite{Bardeen:1983qw,Brandenberger:1983vj,Mukhanov:1990me}, and one generally can choose whether the perturbation lives in the conformal factor or in the stress-energy tensor. However, we have shown here that the simplified picture, where $H$, and hence $K$, are constant throughout space with a local (and arbitrarily large) $\delta \rho$ is not a solution to Einstein's equations for large values of $\zeta$.  The only way to to solve the Hamiltonian constraint in this case is to introduce a spatially varying $K$ to reduce the effective energy density in Eq.~\ref{ham2}. On the other hand, if one chooses to identify $\zeta$ with the conformal factor, while keeping $\delta \rho/H$ small, then one might end up studying the strong branch. 

Similarly, these results are interesting in the context of studies of initial data for inflationary scenarios. Balancing large inhomogeneities with a large extrinsic curvature (effectively increasing the initial expansion) might seem like a biased thing to do in the context of studying whether inflation can get started in the presence of inhomogeneities, but in the case we describe here it may be the only choice for which physical solutions exist for large overdensities. One caveat is that here we match to an asymptotically spatially flat FLRW universe, but we expect that similar considerations would apply if the scenario were to be generalised. Constraints on the existence of solutions gives us useful information about the space of possible initial data, and how that space may differ from the perturbative FLRW picture. The result that the corresponding strong and weak solutions have the same entropy (as approximated by the proper area of their apparent horizons) is also potentially interesting, since arguments about relative entropies are sometimes invoked to decide which initial data are ``more likely'' \cite{Penrose:1988mg, Carroll:2004pn, Albrecht:2004ke,Garfinkle:2023vzf} in this context.

Finally, we note that the non-existence of solutions for some choices of $\delta \rho$, as well as the non-uniqueness of solutions if they do exist, is a direct consequence of the non-linearity of the Hamiltonian constraint.  This nonlinear effect is scale-dependent: the larger the size of the over-dense region the tighter the constraint is on the size of the effective energy density, $\delta \rho - \delta K\left(2+\delta K\right)$.  Evidently, a perturbative treatment of the problem, in which the Hamiltonian constraint is linearized, would completely miss these issues.

\section*{Acknowledgments}
It is a pleasure to acknowledge the Axions in Stockholm workshop as well as the Undergraduate Cosmology Workshop at MIT, where conversations leading up to this work began.  We would like to thank Josu Aurrekoetxea, Sam E. Brady, Anne Christine Davis, Mary Gerhardinger, Alan Guth, David Kaiser, Aron Kovacs and Amanda Miller for helpful conversations.  KC acknowledges support from the Simons Foundation International and the Simons Foundation through Simons Foundation grant SFI-MPS-BH-00012593-03, a UKRI Ernest Rutherford Fellowship (grant number ST/V003240/1) and an STFC Research Grant ST/X000931/1 (Astronomy at Queen Mary 2023-2026). This work was supported in parts by National Science Foundation (NSF) grant PHY-2308821 to Bowdoin College, and PHY-2309919 to Kenyon College.

\bibliography{mybib}

\end{document}